\documentclass[referee]{raa}
\usepackage{graphicx,times}
\usepackage{natbib}
\usepackage{threeparttable}
\usepackage{rotating}
\usepackage{adjustbox}
\usepackage{amsmath,graphicx}

\begin{document}
\title{Evolutions of CH$_3$CN abundance in molecular clumps}

\author{Zhen-Zhen He\inst{1}, Guang-Xing Li\inst{2}, Chao Zhang\inst{1}}

\institute{Department of Astronomy, Yunnan University, Kunming, 650091, China {\it he\_zhenzhen@foxmail.com}
  \and South-Western Institute For Astronomy Research, Yunnan University, Kunming, 650091, China {\it gxli@ynu.edu.cn}}

\abstract{
To investigate the effects of massive star evolution on surrounding molecules, we select 9 massive clumps previously observed with the Atacama Pathfinder Experiment (APEX) telescope and the Submillimeter Array (SMA) telescope.
Based on the observations of APEX, we obtain luminosity to mass ratio L$_{\rm clump}$/M$_{\rm clump}$ that range from 10 to 154 L$_{\sun}$/M$_{\sun}$, where some of them embedded Ultra Compact (UC) H\,{\footnotesize II} region.
Using the SMA, CH$_3$CN (12$_{\rm K}$--11$_{\rm K}$) transitions were observed toward 9 massive star-forming regions.
We derive the CH$_3$CN rotational temperature and column density using XCLASS program, and calculate its fractional abundance.
We find that CH$_3$CN temperature seems to increase with the increase of L$_{\rm clump}$/M$_{\rm clump}$ when the ratio is between 10 to 40 L$_{\sun}$/M$_{\sun}$, then decrease when L$_{\rm clump}$/M$_{\rm clump}$ $\ge$ 40 L$_{\sun}$/M$_{\sun}$.
Assuming the CH$_3$CN gas is heated by radiation from the central star, the effective distance of CH$_3$CN relative to the central star is estimated.
The distance range from $\sim$ 0.003 to $\sim$ 0.083 pc, which accounts for $\sim$ 1/100 to $\sim$ 1/1000 of clump size.
The effective distance increases slightly as L$_{\rm clump}$/M$_{\rm clump}$ increases (R$_{\rm eff}$ $\sim$ (L$_{\rm clump}$/M$_{\rm clump}$)$^{0.5\pm0.2}$).
Overall, the CH$_3$CN abundance is found to decrease as the clumps evolve, e.g., X$_{\rm CH_3CN}$ $\sim$ (L$_{\rm clump}$/M$_{\rm clump}$)$^{-1.0\pm0.7}$.
The steady decline of CH$_3$CN abundance as the clumps evolution can be interpreted as a result of photodissociation.
}

\maketitle

\section{Introduction} \label{sec:intro}

Massive stars (M $\geq$ 8 M$_{\sun}$) are formed inside molecular clouds.
Feedback from star formation has substantial impacts on the evolution of the surround interstellar medium (ISM) through outflows, winds, as well as strong UV radiation \citep{1999PASP..111.1049G,2007ARA&A..45..481Z}.
Molecules, which have been detected in the ISM or circumstellar shells, are powerful tools to probe physical conditions such as densities, gas temperatures, and kinematical properties \citep{2009ARA&A..47..427H}.

\citet{2014A&A...563A..97G} observed a sample of high-mass star-forming regions at different evolutionary stages with the IRAM 30m telescope, those stages varying from infrared dark clouds (IRDCs) to high-mass protostellar objects (HMPOs) to hot molecular cores (HMCs) and, finally, ultra-compact (UC) H\,{\footnotesize II} regions.
They found that molecular species are rich in the HMC phase and decline for the UC H\,{\footnotesize II} stage.
However, the evolution of high-mass star formation occurs on a short time scale and in a clustered environment \citep{2003MNRAS.343..413B}. The transition from one stage to the next is not well-defined, there are some overlaps among those stages.

The collapse of molecular clump\footnote{Following \citet{2009ApJ...696..268Z,2015ApJ...804..141Z} in this paper, we tend to refer the size of cloud as 10 - 100 pc, the size of molecular clump as $\leq$ 1 pc, and the size of dense core as 0.01 - 0.1 pc.} is a global process: although the star formation that we observe occurs mostly at the center of the clumps, the actual collapse occurs on a much larger scale \citep[e.g.,][]{2015ApJ...804...44M,2018MNRAS.477.4951L}.
To study this collapse and the associated star formation, it is necessary to correlate the tracers of star formation with the properties of their natal clumps.
A promising evolutionary indicator for massive and dense clumps is the L/M ratio -- the ratio between the bolometric luminosity and the mass of the clump.
Single dish investigations about the physical properties of star-forming regions at different L/M have been conducted in some works using molecules as probes \citep[e.g., CH$_3$C$_2$H, CH$_3$CN, NH$_3$, CH$_3$OH;][]{2016ApJ...826L...8M,2017A&A...603A..33G}.
These works show that the molecular temperatures are significantly changed in different physical conditions during stars formation.

CH$_3$CN (methyl cyanide) is known as a good tracer of warm and dense gas \citep[e.g.,][]{2001ApJ...558..194P,2005ApJS..157..279A,2017A&A...603A..33G}.
High spatial resolution observations seem to indicate that this molecule traces the disks around forming stars \citep{1997A&A...325..725C,2010MNRAS.406..102K,2014A&A...566A..73C,2016ApJ...823..125C,2019A&A...623A..77S}.
Interferometric observations with high spatial resolutions are suitable for studying small regions as the CH$_3$CN emission mostly originates from small scales.
\citet{2014ApJ...786...38H} searched the literature for MSFRs in the HMC phase observed using Submillimeter Array (SMA) and selected 17 sources.
They studied the chemical properties of CH$_3$CN, and suggested that CH$_3$CN was formed in gas phase and their abundance was increased with the increase of temperature.
In this paper, we tend to use the sources selected by \citet{2014ApJ...786...38H} and place these sources in the context of star evolution to study the influence of star formation on the surrounding molecules.
Following \citet{2008A&A...481..345M}, we propose to study star evolutional stages indicated by the luminosity to mass ratios of the clumps.

\section{Data} \label{sec:data}

\subsection{Sample}
Our samples are taken from \citet{2014ApJ...786...38H}, who searched the literature for MSFRs in the HMC phase.
Most of their sample sources are associated with UC H\,{\footnotesize II} regions.
These sources were observed at a spatial scale of $\sim$0.1 pc with similar observational resolution.
We matched them with Atacama Pathfinder Experiment (APEX) Telescope Large Area Survey of the Galaxy \citep[ATLASGAL;][]{2009A&A...504..415S} dense clump catalogue based on their positions.
The survey aims to provide a large and systematic list of clumps in the early embedded evolution stage.
Nine massive clumps G5.89--0.39, G8.68--0.37, G10.62--0.38, G28.20--0.04N, G45.07+0.13, G45.47+0.05, IRAS 16547--4247, IRAS 18182--1433 and IRAS 18566+0408 are selected after removing some non-matched sources.
The clump distances, effective radii, masses, and bolometric luminosities are taken from \citet{2018MNRAS.473.1059U}.
For our purpose, we use L$_{\rm clump}$/M$_{\rm clump}$ -- the ratio between clump luminosity and mass -- as the evolutionary indicator, and the 9 targets have L$_{\rm clump}$/M$_{\rm clump}$ ranging from 10 to 154 L$_{\sun}$/M$_{\sun}$.
A summary of clump properties is presented in Table \ref{tab:clump}.

\begin{table}
\begin{minipage}[]{100mm}
\caption{Clump properties\label{tab:clump}}\end{minipage}
\setlength{\tabcolsep}{1pt}
\begin{tabular}{lccccc}
\hline\noalign{\smallskip}
Source & Distance & Mass      & Luminosity          & L$_{\rm clump}$/M$_{\rm clump}$ & R$_{\rm clump}$ \\
       & kpc      & M$_{\sun}$& 10$^{5}$ L$_{\sun}$ & L$_{\sun}$/M$_{\sun}$           & pc \\
\hline\noalign{\smallskip}
G5.89--0.39      & 2.99 & 1698 & 2.11 & 124 & 0.653 \\
G8.68--0.37      & 4.45 & 2218 & 0.23 & 10  & 0.712 \\
G10.62--0.38     & 4.95 & 8649 & 5.21 & 60  & 1.584 \\
G28.20--0.04N    & 6.05 & 4446 & 1.30 & 29  & 1.349 \\
G45.07+0.13      & 8.00 & 3111 & 4.82 & 154 & 1.357 \\
G45.47+0.05      & 8.40 & 7161 & 4.30 & 59  & 2.078 \\
IRAS 16547--4247 & 2.74 & 1678 & 0.60 & 35  & 0.691 \\
IRAS 18182--1433 & 4.71 & 1342 & 0.18 & 13  & 0.708 \\
IRAS 18566+0408  & 4.86 & 1940 & 0.23 & 11  & 1.084 \\
\noalign{\smallskip}\hline
\end{tabular}
\tablecomments{\textwidth}{Properties of our sample clumps taken by matching them to the sample of ATLASGAL \citep{2018MNRAS.473.1059U}.}
\end{table}

\subsection{SMA observations and data reduction}
The SMA observations were performed between April 2004 and January 2009.
Selected targets are listed in Table \ref{tab:ObsPara}, where observational epoch, frequency range, and spectral resolution are included.
The correlator was set to the double-sideband receiver --- the lower sideband (LSB) and the upper sideband (USB), each bandwidth is about 2 GHz.
The rest frequencies of CH$_3$CN (12$_{\rm K}$--11$_{\rm K}$) are covered in the lower sideband.
Spectral resolution is 0.406 MHz ($\sim$0.53 km s$^{-1}$) or 0.812 MHz ($\sim$1.1 km s$^{-1}$) for different sources.
The system temperatures during those observations were less than 600 K.

Calibration and imaging are performed using the MIRIAD package\footnote{http://www.cfa.harvard.edu/sma/miriad} \citep{1995ASPC...77..433S}.
Calibrators of bandpass, gain, and flux for each target are listed in Table \ref{tab:ObsPara}.
Considering the SMA monitoring of quasars, we estimate the fluxes uncertainty of $\sim$20\%.
We combined the continuum data from LSB and USB, the synthesized beam sizes range from 1.83\arcsec $\times$ 0.96$''$ to 5.37\arcsec $\times$ 2.66\arcsec.
All the line data was regrided to a uniform spectral resolution of 0.812 MHz ($\sim$1.1 km s$^{-1}$).

\begin{figure*}
\centering
\includegraphics[width=\textwidth]{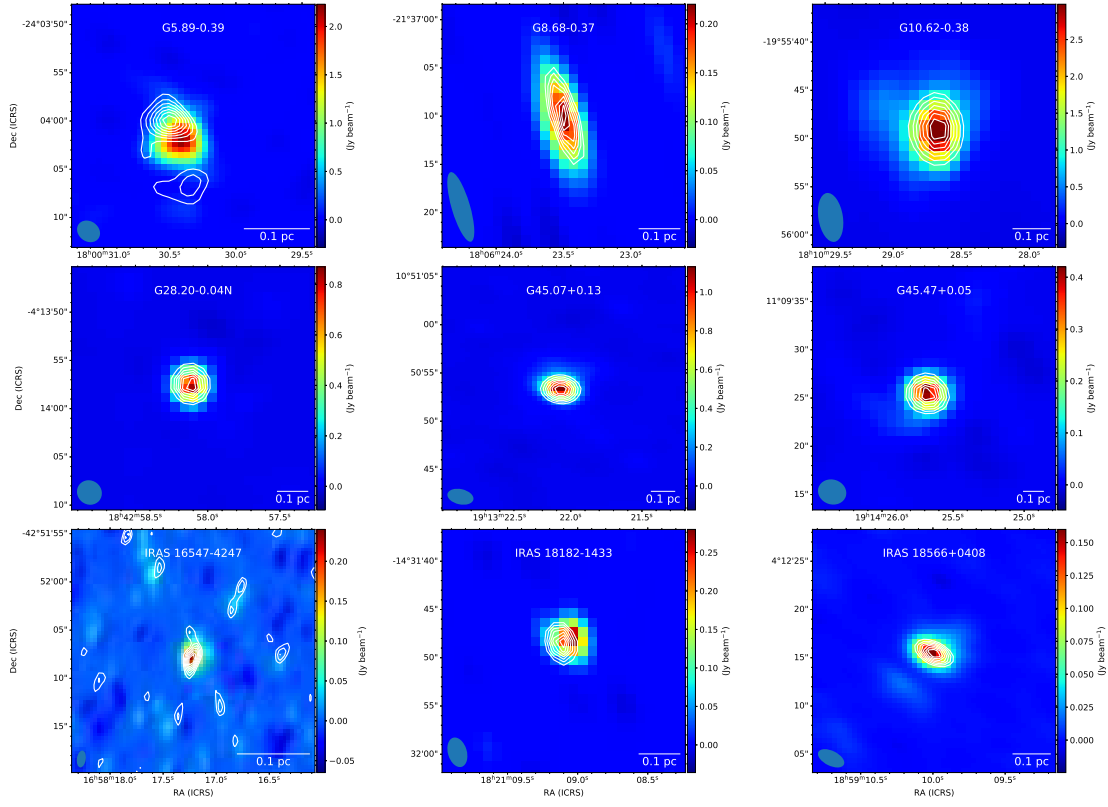}
\caption{1.3 mm continuum images (color scale) and molecules line emission (white contours) of CH$_3$CN (12$_2$--11$_2$) toward massive star-forming regions. Contour levels have steps of 30\%, 40\%, 50\%, 60\%, 70\%, 80\%, until 90\% of integrated emission. The synthesized beam of each source is shown at the bottom left. The 1.3 mm integrated emission flux of G5.89--0.39, G8.68--0.37, G10.62--0.38, G28.20--0.04N, G45.07+0.13, G45.47+0.05, IRAS 16547--4247, IRAS 18182--1433 and IRAS 18566+0408 is 7.50 Jy, 0.38 Jy, 6.86 Jy, 1.15 Jy, 1.63 Jy, 0.60 Jy, 0.56 Jy, 0.47 Jy, and 0.31 Jy, respectively, which are listed in Table \ref{tab:tab}.}
\label{fig:ContinuumImages}
\end{figure*}

\section{results and Analysis} \label{sec:results}
\citet{2014ApJ...786...38H} had completed the analysis of those data, they derived CH$_3$CN temperature and column density using a two-component method --- a hot dense component and a warm extended component --- by assuming different source size.
In their results, the molecular abundances are mainly contributed from the hot dense components.
In order to study the global variation of molecular abundance in star formation, we reanalysised selected data using single component fitting, the source size was derived by fitting CH$_3$CN (12$_{\rm 2}$--11$_{\rm 2}$) line emission region of each target (see Section \ref{sec:MolLine}), this method has been used in previous analysis \citep{2006A&A...454..221B}.

\subsection{Continuum} \label{sec:cont}
The continuum images, as shown in Figure \ref{fig:ContinuumImages}, are constructed from line-free channels of LSB and USB in the visibility domain.
The 1$\sigma$ noise level of continuum is lower than 20 mJy beam$^{-1}$.
Two-dimensional gaussian fittings are performed to obtain peak intensities, total flux densities, and deconvolved source sizes.
The relevant results are listed in Table \ref{tab:tab}.

The 1.3 mm continuum emission may be contributed by free-free radiation and dust thermal radiation, because most HMCs harbor embedded UC H\,{\footnotesize II} regions.
The free-free emission was obtained from \citet{2014ApJ...786...38H}, who assumed the free-free emission is optically thin from 10 to 100 GHz and estimated the free-free emission by $S_{\nu}\propto \nu^{-0.1}$.
The estimated dust continuum flux of each source is listed in Table \ref{tab:tab}.


To derive CH$_3$CN fractional abundance, we first calculate H$_2$ column densities for all observed regions.
Assuming that the dust emission is optically thin, H$_2$ column density can be obtained by \citep{1983QJRAS..24..267H,1991ApJ...380..429L}:
\begin{equation}
N_{\rm H_2} = 8.1\times10^{17} \frac{{e^{{h\nu}/k T}}-1}{{Q(\nu)}\Omega} \left(\frac{S_\nu}{\rm Jy}\right) \left(\frac{\nu}{\rm GHz}\right)^{-3}\left(\rm cm^{-2}\right),
\end{equation}
where $k$ and $h$ are the Boltzmann constant and Planck constant, respectively, $T$ is the dust temperature, Q($\nu$) is the grain emissivity at frequency $\nu$, $\Omega$ is the solid angle of source, S$_\nu$ is the total integrated flux of the dust continuum.
We adopt Q($\nu$) = 2.2 $\times$ 10$^{-5}$ at 1.3 mm \citep[$\beta$ = 1.5;][]{1990ApJ...356..195L,1991ApJ...380..429L}, and the gas-to-dust ratio of 100 is used.
The dust temperature can be estimated from CH$_3$CN rotational temperature, which will be derived in Section \ref{sec:MolLine}, assuming that dust and gas are in thermal equilibrium \citep{1998ApJ...497..276K}.
The derived H$_2$ column densities range from 0.39 $\times$ 10$^{24}$ to 4.73 $\times$ 10$^{24}$ cm$^{-2}$, the values are listed in Table \ref{tab:tab}.

Considering an absolute flux uncertainty of $\sim$20\% for SMA observations, dust temperature uncertainty of $\sim$20\%, we estimate H$_2$ column density uncertainties of $\sim$50\% \citep{2019ApJ...886..102S}.
The H$_2$ column density we derived is comparable to the result of \citet{2014ApJ...786...38H}.
The difference may be caused by that they used the CH$_3$CN temperature of hot dense component to assume the dust temperature, while we derived a relatively low CH$_3$CN temperature using single component fitting.
The H$_2$ column density is sensitive to dust temperature.

\begin{figure*}
\centering
\includegraphics[width=.33\textwidth]{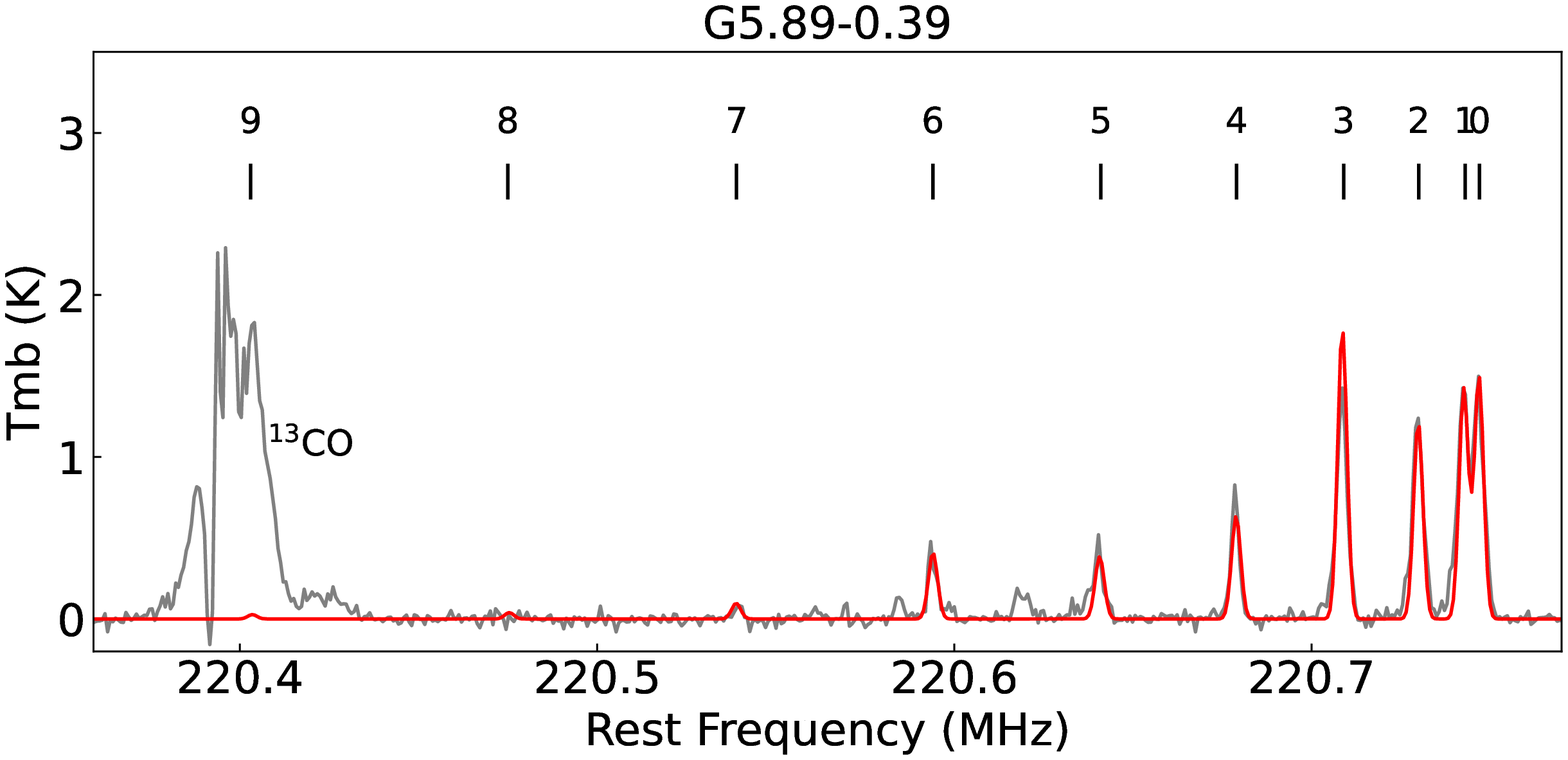}
\includegraphics[width=.33\textwidth]{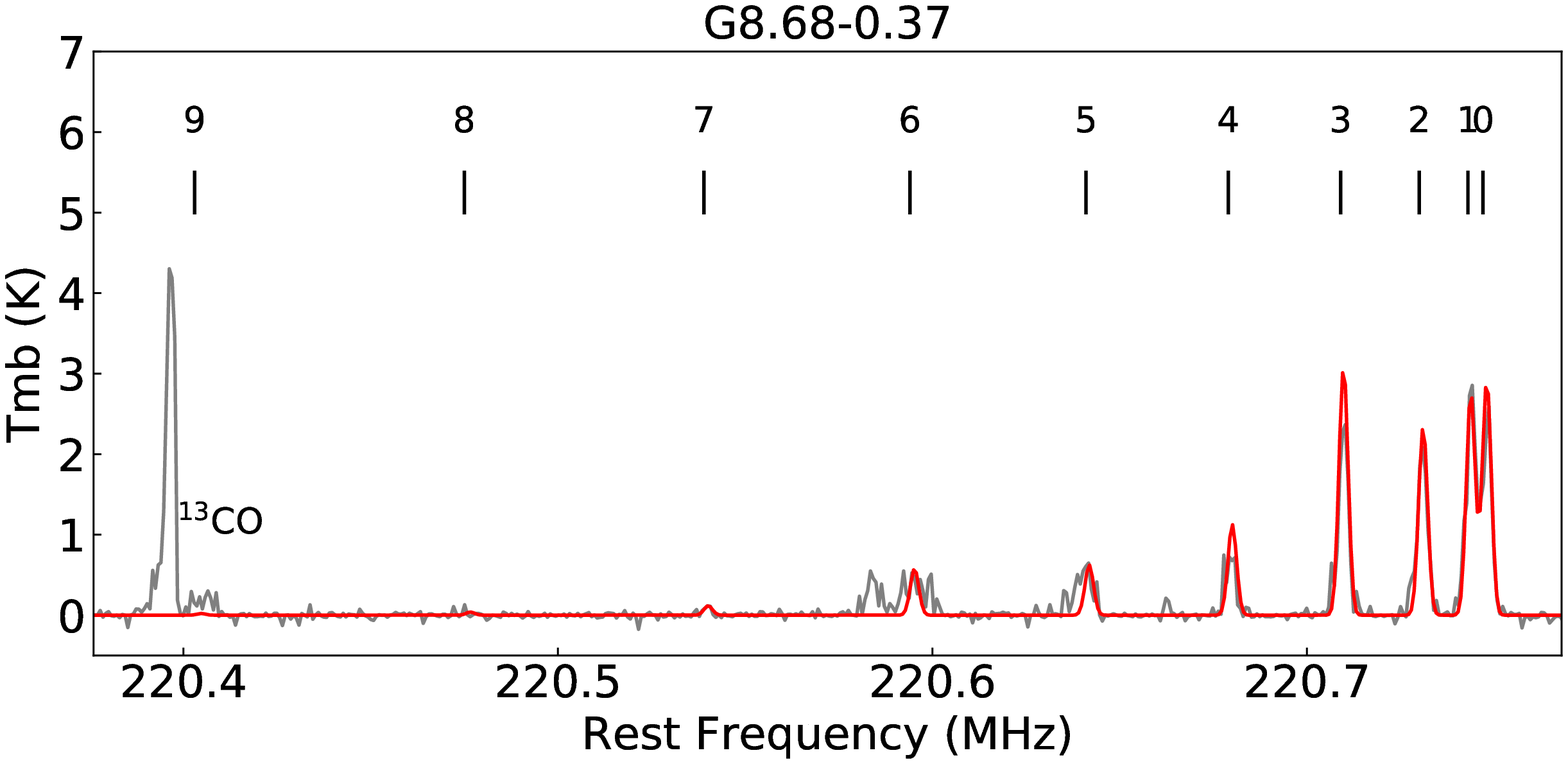}
\includegraphics[width=.33\textwidth]{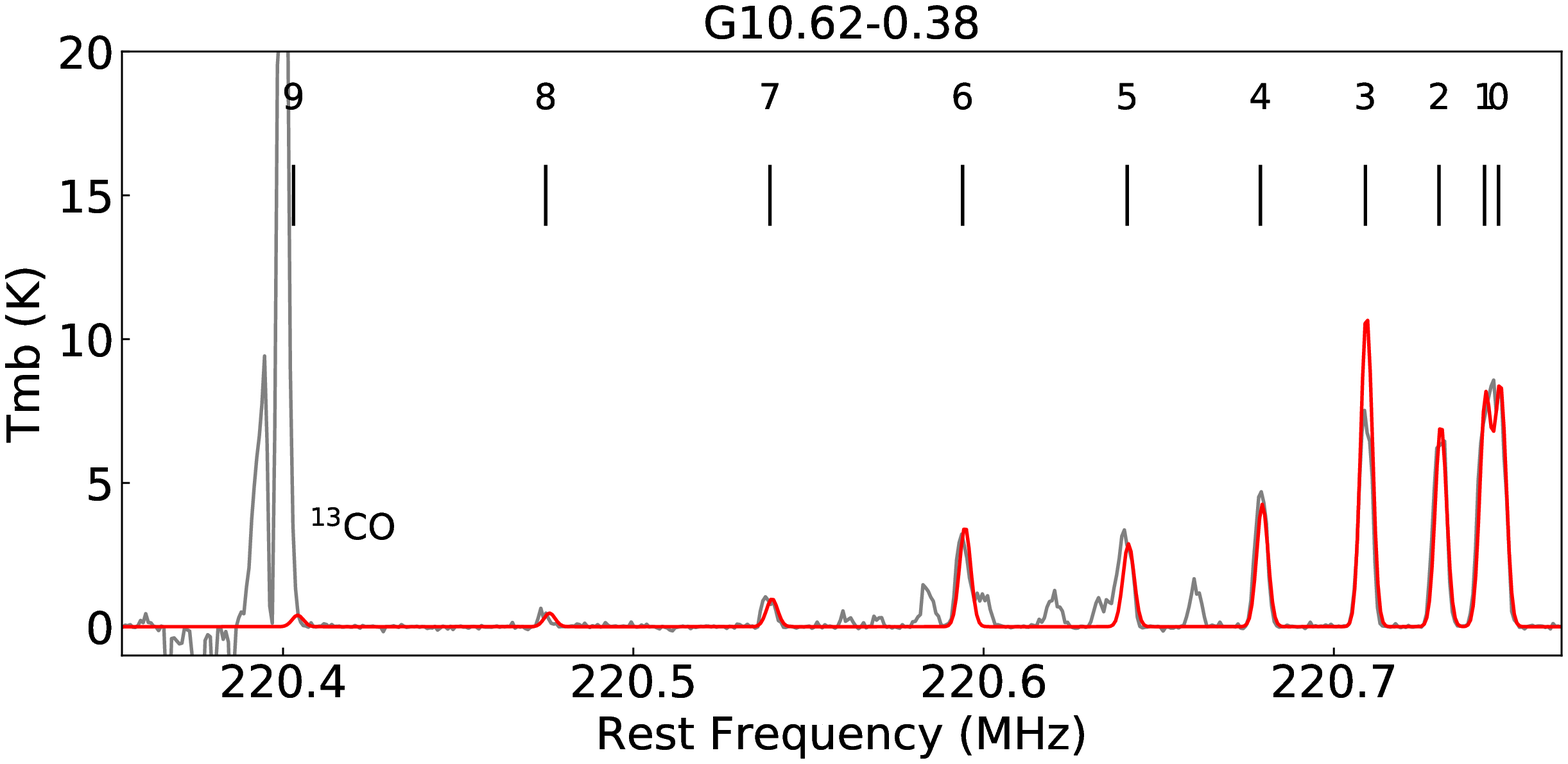}\\
\includegraphics[width=.33\textwidth]{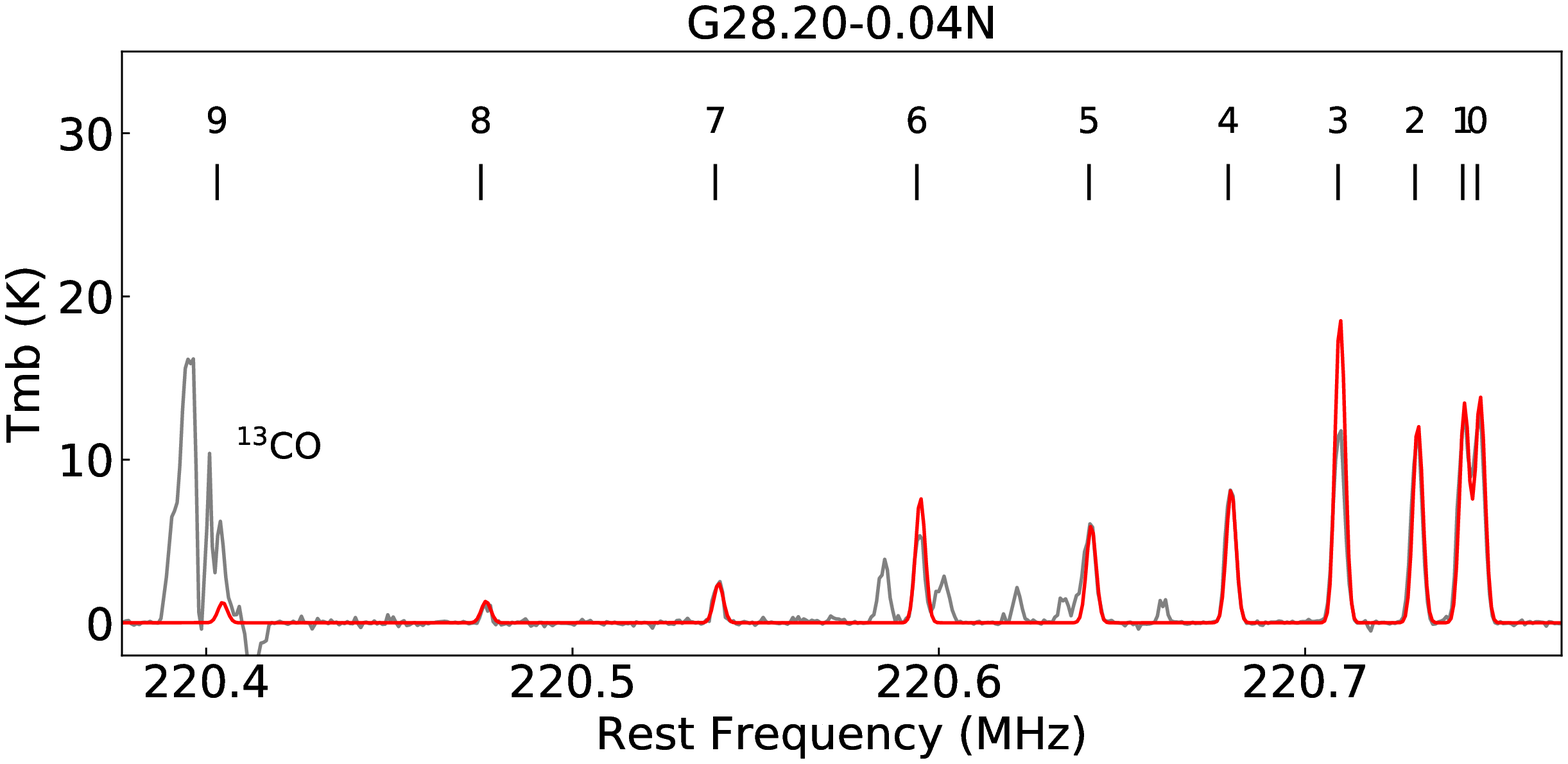}
\includegraphics[width=.33\textwidth]{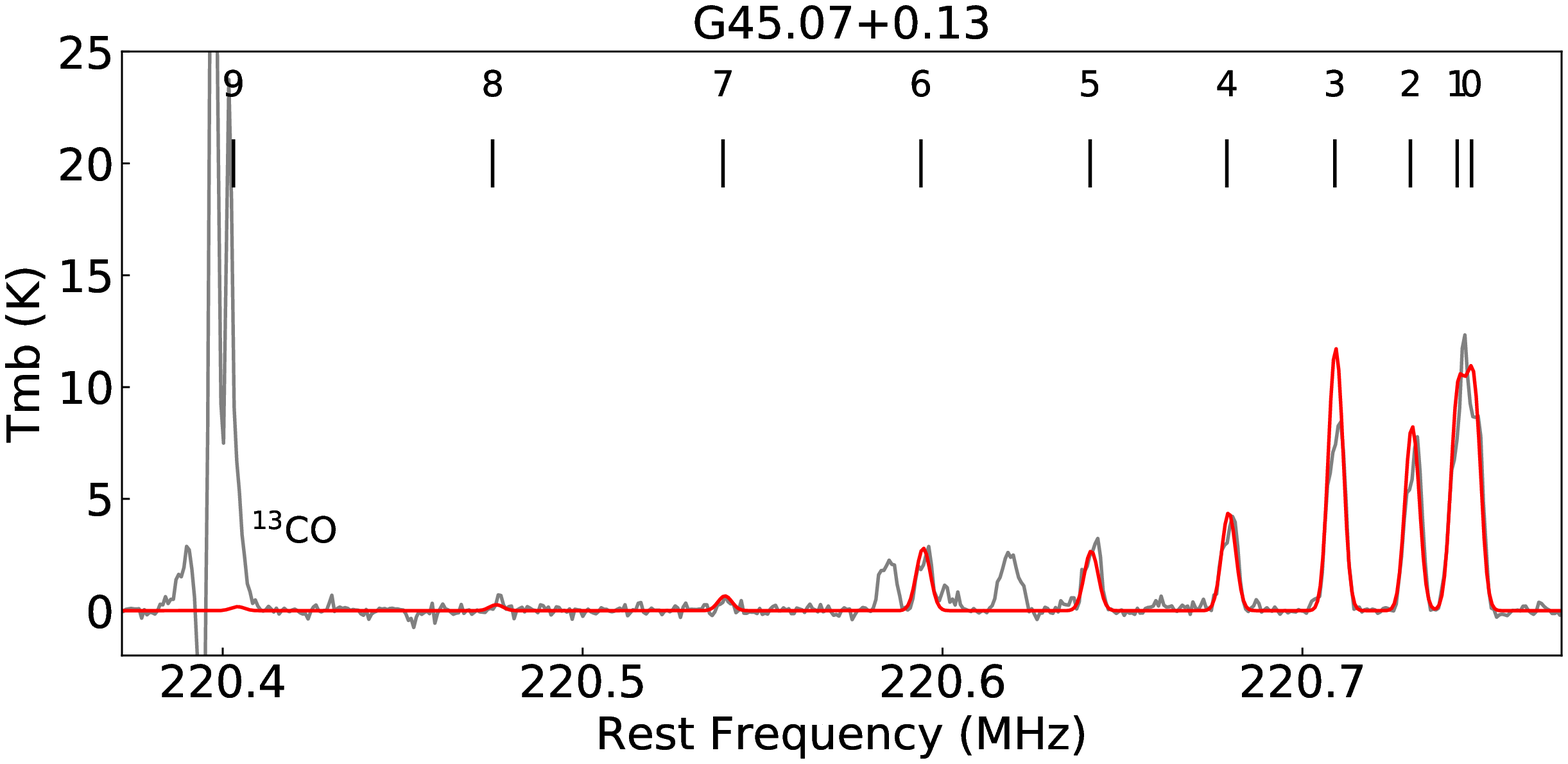}
\includegraphics[width=.33\textwidth]{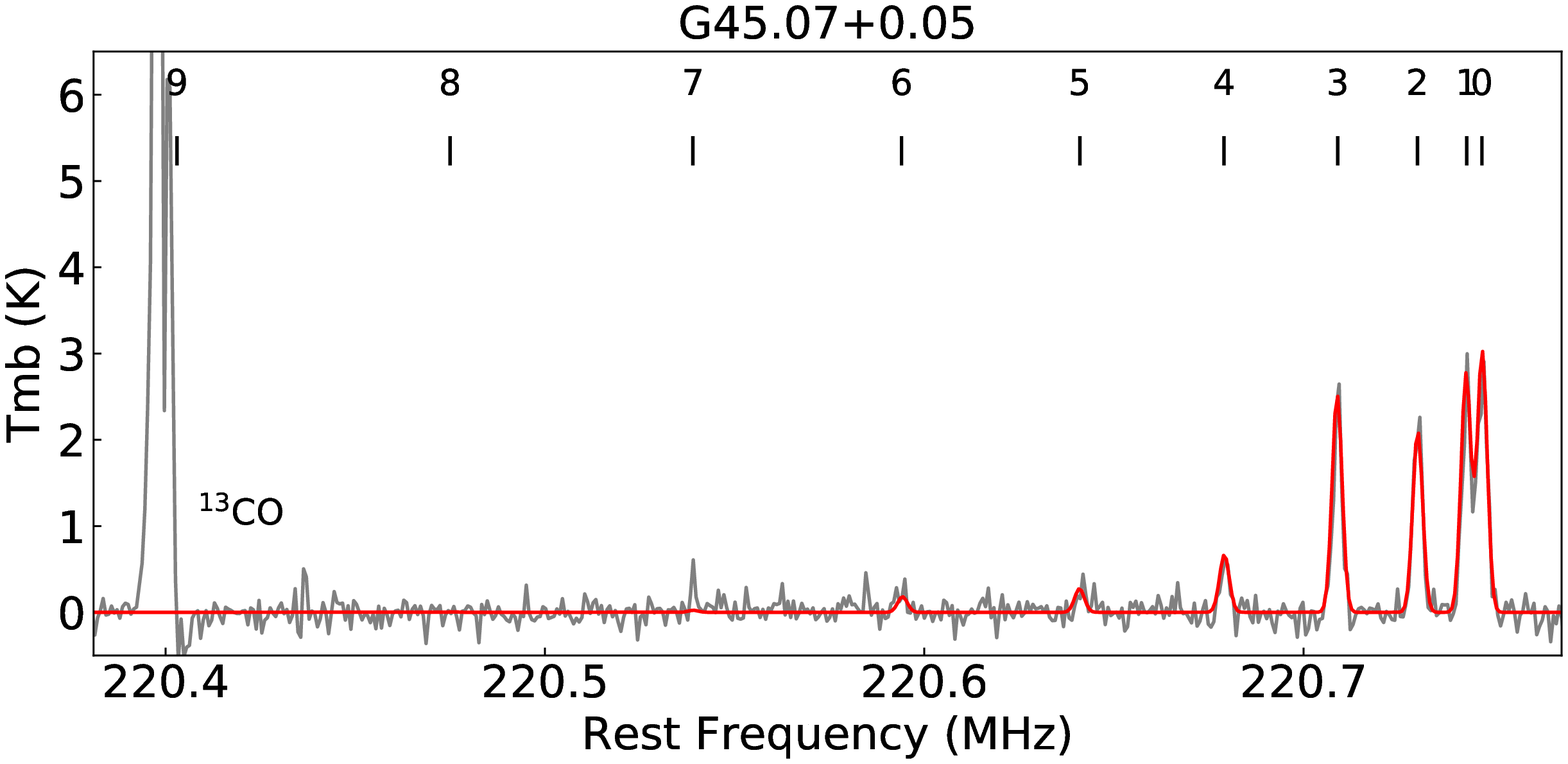}\\
\includegraphics[width=.33\textwidth]{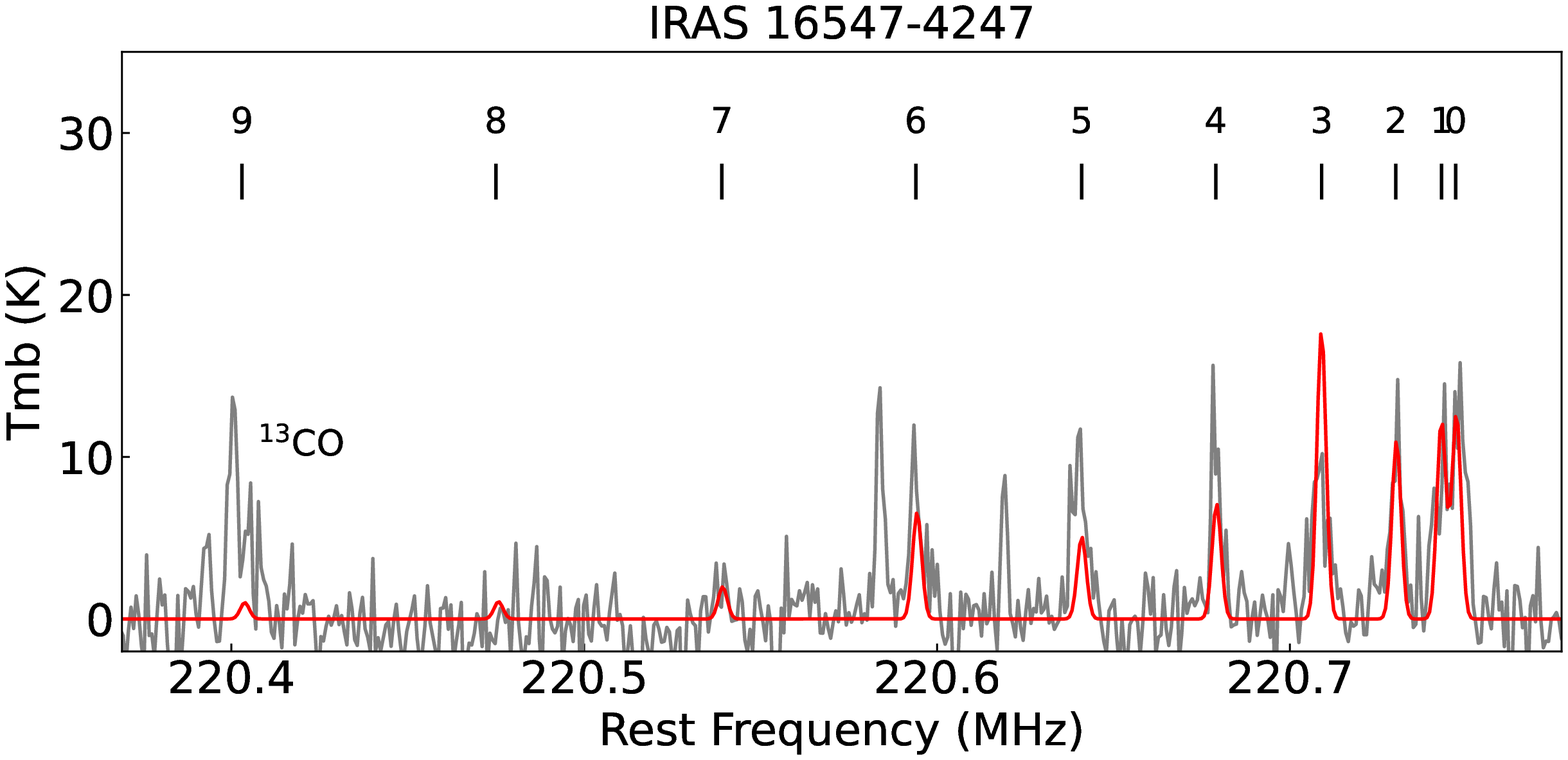}
\includegraphics[width=.33\textwidth]{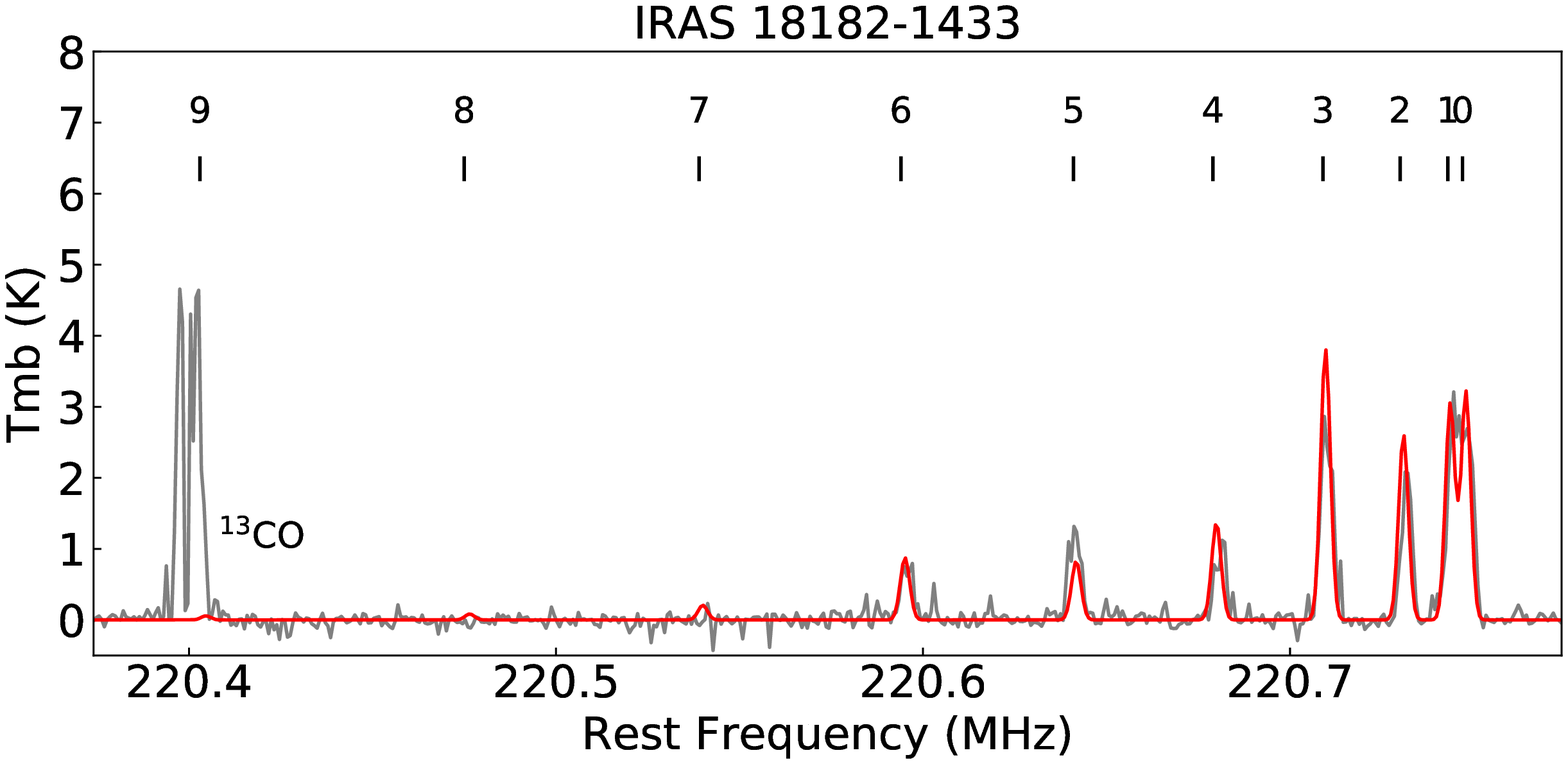}
\includegraphics[width=.33\textwidth]{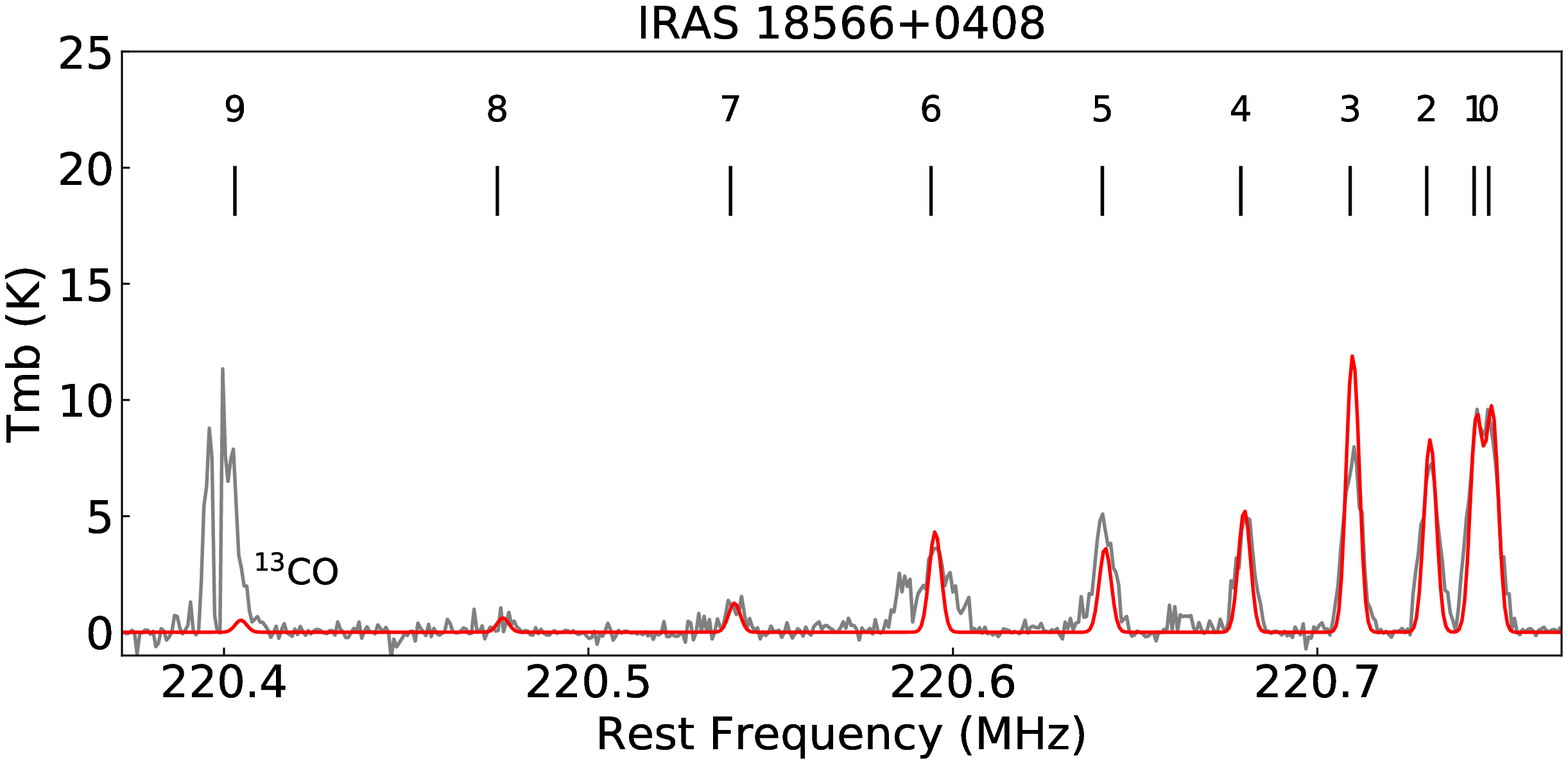}\\
\caption{Observed spectra (black) and model spectra (red) of CH$_3$CN (12$_{\rm K}$--11$_{\rm K}$). The model spectra are derived with XCLASS package. The number represent the $K$-ladder quantum numbers. The $K$ = 9 line is overlapped with $^{13}$CO (2-1) in some regions.}
\label{fig:linedata}
\end{figure*}

\subsection{Molecular line emission}\label{sec:MolLine}
The spectral lines are extracted from the continuum peak position.
Their continuum-subtracted spectra are plotted with intensity in units of Kelvin in Figure \ref{fig:linedata}.
The CH$_3$CN (12$_{\rm K}$--11$_{\rm K}$) lines are identified following \citet{2014ApJ...786...38H}.

Spectral lines are modelled using the XCLASS package\footnote{http://xclass.astro.uni-koeln.de} \citep{2017A&A...598A...7M} to derive molecular temperature and column density.
The modeling parameters are source size, rotational temperature, column density, line width, and velocity offset.
The source sizes are obtained by two-dimension Gaussian fits to CH$_3$CN (12$_2$--11$_2$) line images.
The 1$\sigma$ noise level in line images are lower than 100 mJy beam$^{-1}$.
The line width and velocity offsets are obtained by Gaussian fitting.
Those parameters are used as initial parameters.
The optimization algorithm in MAGIX is used to explore the parameter space and minimize the $\chi^2$ distribution space.
The detailed fitting functions and modelling procedures are described in \citet{2017A&A...598A...7M}, where local thermodynamical equilibrium (LTE) is assumed.
The derived molecular temperature and column density are presented in Table \ref{tab:tab}.

We derive molecular fractional abundances relative to H$_2$ using X = N$_{\rm tot}$/N$_{\rm H_2}$, where N$_{\rm H_2}$ is the H$_2$ column density.
CH$_3$CN rotational temperatures, column densities, and abundances are listed in Table \ref{tab:tab}.
Both temperature and column density of CH$_3$CN obtained by our single component fitting are lower than yet compareable to that of the hot dense component and higher than that of the warm extended component obtained by \citet{2014ApJ...786...38H}.

\section{Discussion} \label{sec:Dis}

\subsection{Evolutionary stages}
\begin{figure}
\centering
\includegraphics[width=.6\textwidth]{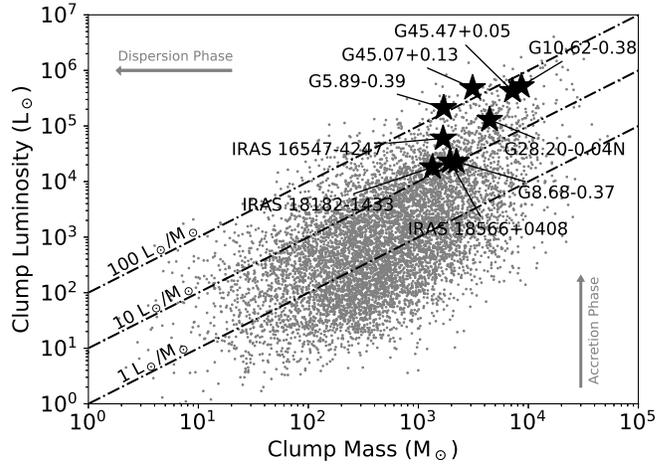}
\caption{Distribution of sources in the luminosity-mass plane. Stars represent selected sources in this paper. Gray dots represent the ATLASGAL clumps from \citet{2018MNRAS.473.1059U}. The lower, middle and upper diagonal dash-dot lines indicate the L$_{\rm clump}$/M$_{\rm clump}$ = 1, 10 and 100 L$_{\sun}$/M$_{\sun}$, respectively. Vertical and horizontal arrow refer to as the accretion and the envelope dispersion phases derived by \citet{2008A&A...481..345M}.}
\label{fig:pair_lm}
\end{figure}

The distribution of ATLASGAL sources in the luminosity-mass (L$_{\rm clump}$-M$_{\rm clump}$) plane are presented in Figure \ref{fig:pair_lm}.
This type of diagram has been used in the studies of low-mass and high-mass star-forming regions \citep[e.g.,][]{1996A&A...309..827S,2008A&A...481..345M,2017A&A...603A..33G}, and can be used as a tool for separating different evolutionary stages.
Evolutionary tracks were derived by \citet{2008A&A...481..345M}, vertical and horizontal arrow refer to as the accretion and the envelope dispersion phases.
Three lines of constant L$_{\rm clump}$/M$_{\rm clump}$ ratios (i.e., 1, 10 and 100 L$_{\sun}$/M$_{\sun}$) are shown in figure.
Our sources at different evolutionary stages are marked in Figure \ref{fig:pair_lm}.

Based on previous studies, G8.68--0.37, IRAS 18182--1433, and IRAS 18566+0408 have no or weak continuum emission at centimeter observations \citep{2011ApJ...726...97L,2006A&A...454..221B,2005ApJ...618..339A}, and they are massive star forming regions at an early evolutionary stage before forming a significant UC H\,{\footnotesize II} region.
Clumps reaching L$_{\rm clump}$/M$_{\rm clump}$ $\sim$ 10 L$_{\sun}$/M$_{\sun}$ are likely to stay in the transition phase between the main accretion phase and the dispersion phase \citep{2017A&A...603A..33G}.
Most sources with L$_{\rm clump}$/M$_{\rm clump}$ $\ga$ 30 L$_{\sun}$/M$_{\sun}$ are harboring UC H\,{\footnotesize II} regions \citep{2009ApJ...704L...5S,1997ApJ...478..283H,1994AAS...184.3012A,1989ApJ...340..265W}, especially G5.89--0.39 and G45.07+0.13 have a L$_{\rm clump}$/M$_{\rm clump}$ in excess of 100 L$_{\sun}$/M$_{\sun}$.
G5.89--0.39 is an expanding shell-like UC H\,{\footnotesize II} region \citep{1989ApJS...69..831W}. Recently, \citet{2019MNRAS.486L..15Z} have presented sensitive CO(J = 3–2) observations that revealed the possible presence of an explosive outflow in G5.89--0.39.
G45.07+0.13 is a pair of spherical UC H\,{\footnotesize II} regions, this region are interpreted as a shell generated by stellar wind \citep{1984ApJ...277..164T}.
A detailed description of each source is given in \citet{2014ApJ...786...38H}.

Our sources have similar masses yet different luminosity-to-mass ratios.
Taking advantage of this, we propose that our sources G8.68-0.37, IRAS 18566+0408, IRAS 18182-1433, G28.20-0.04N, IRAS 16547-4247, G45.47+0.05, G10.62-0.38, G5.89-0.39, and G45.07+0.13 form a sequence with L$_{\rm clump}$/M$_{\rm clump}$ ranging from 10 to 154 L$_{\sun}$/M$_{\sun}$.

\begin{figure*}
\centering
\includegraphics[width=.45\textwidth]{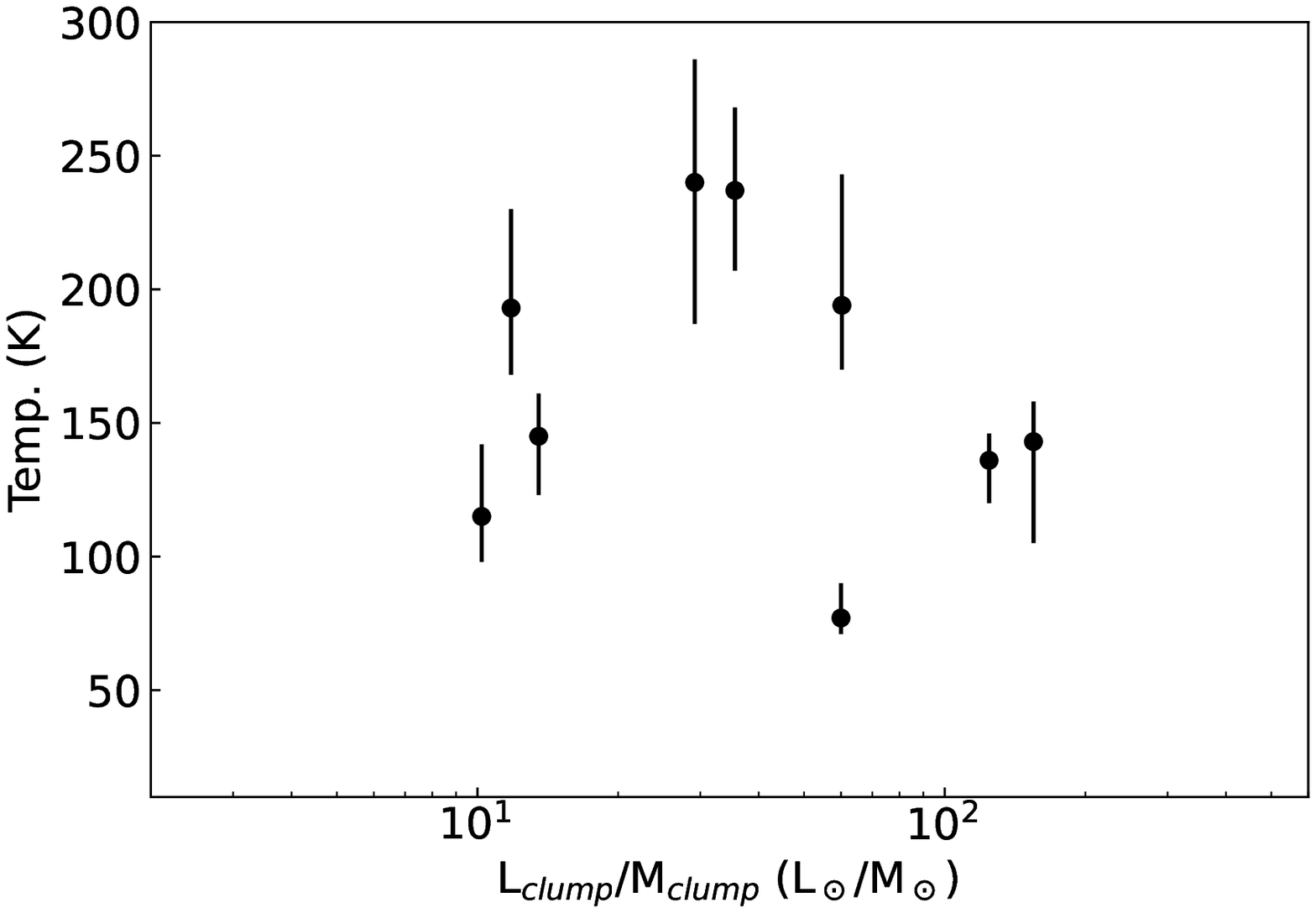}
\includegraphics[width=.45\textwidth]{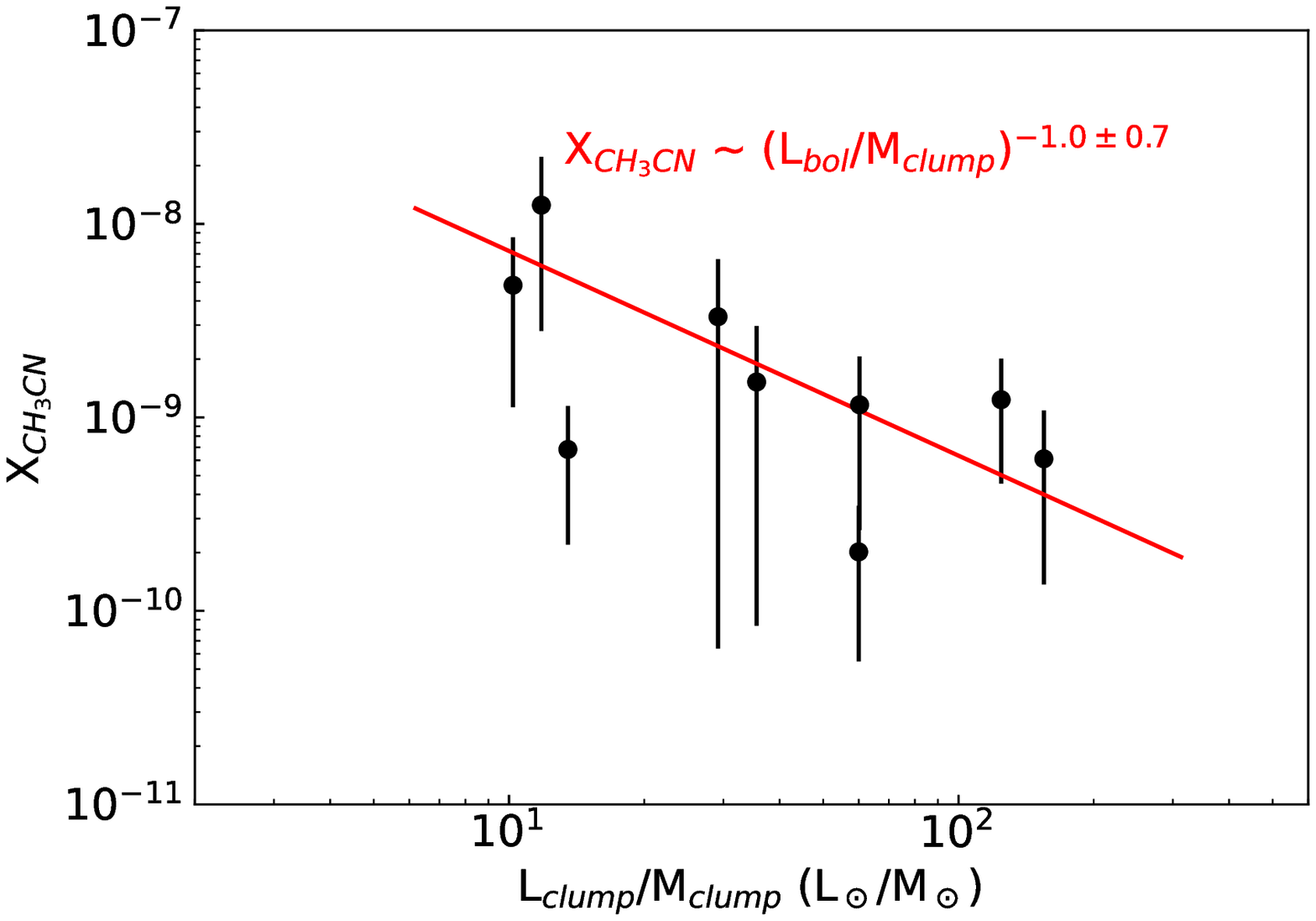}\\
\includegraphics[width=.45\textwidth]{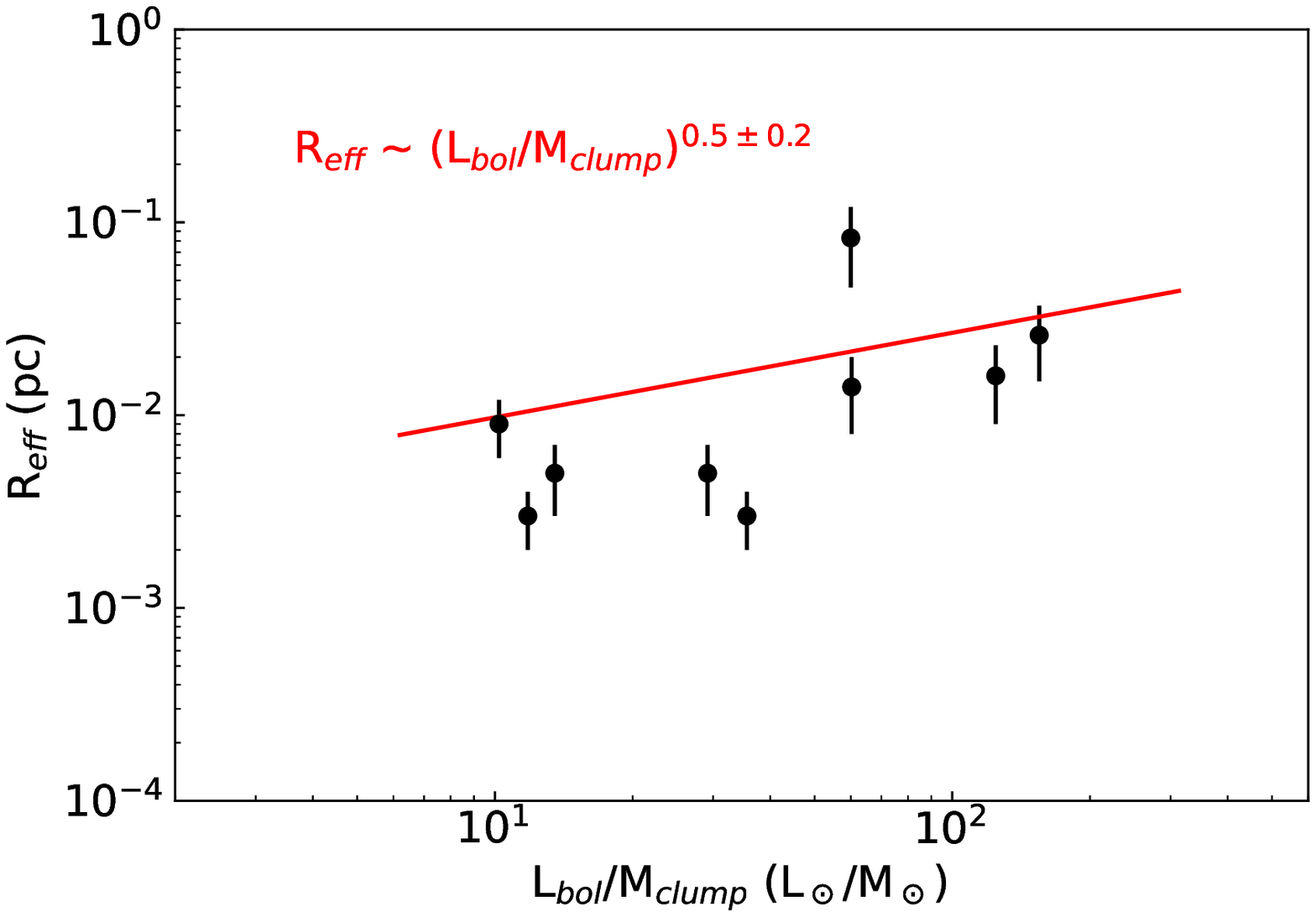}
\includegraphics[width=.45\textwidth]{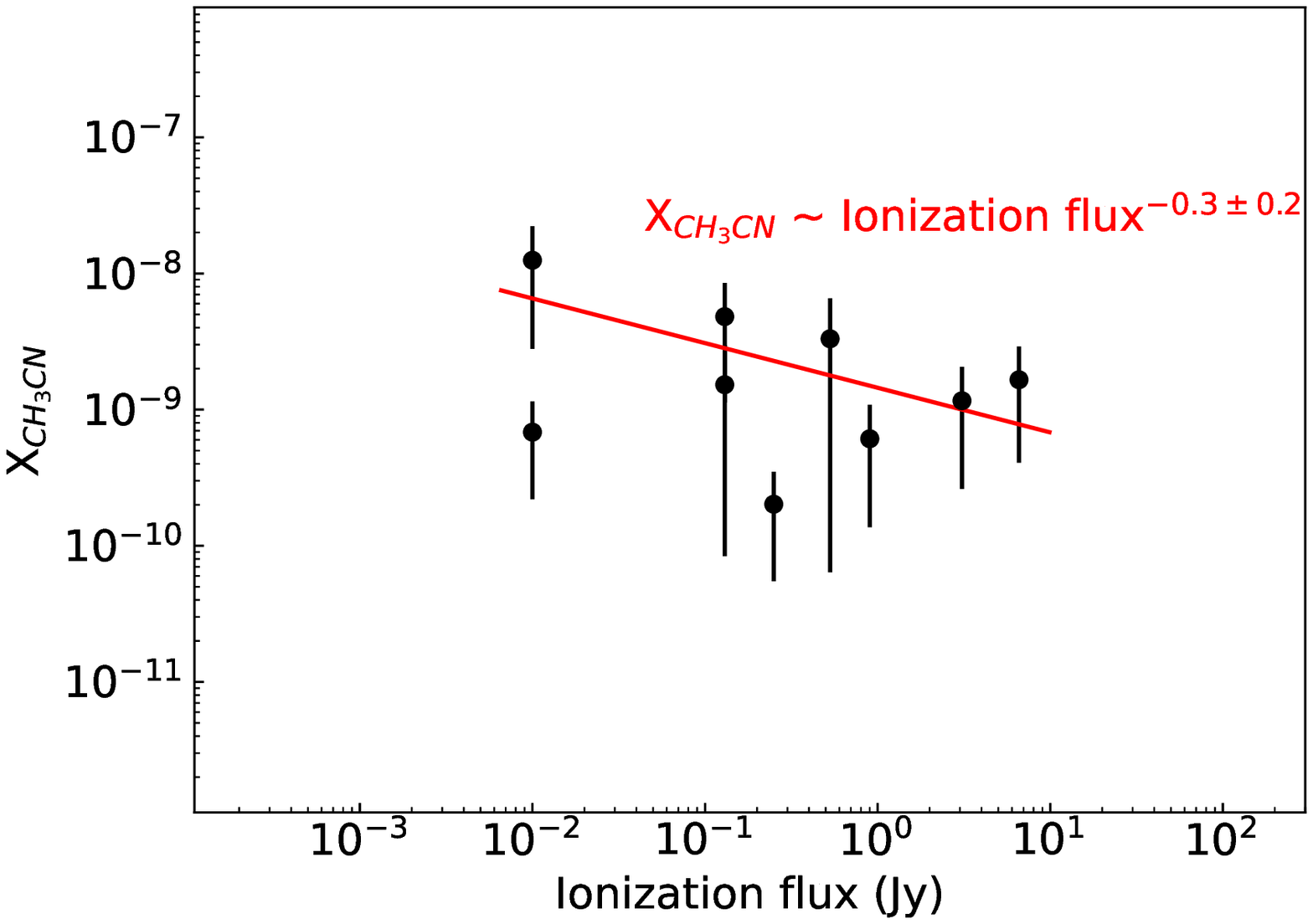}
\caption{{\bf{Upper left:}} CH$_3$CN rotational temperature as a function of the L$_{\rm clump}$/M$_{\rm clump}$. Molecular temperatures are estimated with the XCLASS package. 
{\bf{Upper right:}} CH$_3$CN abundance as a function of the L$_{\rm clump}$/M$_{\rm clump}$, the abundances are derived relative to H$_2$ column density.
{\bf{Lower left:}} Derived size ratios between CH$_3$CN effective distance R$_{\rm eff}$ as a function of the L$_{\rm clump}$/M$_{\rm clump}$.
Molecular effective distance R$_{\rm eff}$ is derived based on molecular temperature and clump bolometric luminosity using $L = 4\pi R^{2}\sigma T^{4}$. 
{\bf{Lower right:}} CH$_3$CN abundance as a function of the ionization flux.
The ionization flux at 1.3 mm was derived by \citet{2014ApJ...786...38H}, who search for reported 10 to 100 GHz emission from the literatures and following $S_{\nu}\propto \nu^{-0.1}$.
 \label{fig:lm}}
\end{figure*}

\subsection{Temperature}
In the upper left panel of Figure \ref{fig:lm}, we show the rotational temperature traced by CH$_3$CN gas as a function of the L$_{\rm clump}$/M$_{\rm clump}$.
It seems that the temperature increases with L$_{\rm clump}$/M$_{\rm clump}$ when the ratio between 10 to 40 L$_{\sun}$/M$_{\sun}$, and decreases when L$_{\rm clump}$/M$_{\rm clump}$ is greater than 40 L$_{\sun}$/M$_{\sun}$.
It is puzzling that rotational temperatures do not demonstrate a monotonic increase over the increase of L$_{\sun}$/M$_{\sun}$ ratios.
One would expect, in an idealized situation, that as a massive protostar gains masses and luminosity, the protostellar heating raises temperatures of the surrounding gas. Therefore, rotational temperatures are expected to increase with L$_{\sun}$/M$_{\sun}$ ratios.
This trend was observed by a single dish observations \citep[e.g.,][]{2016ApJ...826L...8M,2017A&A...603A..33G,2018MNRAS.473.1059U}.
In these literatures, all molecular temperatures are considered to be continuously heated up when L$_{\rm clump}$/M$_{\rm clump}$ $\ga$ 10 L$_{\sun}$/M$_{\sun}$.

Several factors may bias the data.
The most direct influence on the observation results is the spatial resolution. The spatial resolution range of our 9 sources is from $\sim$10000 to $\sim$20000 AU, except for IRAS 16547--4247 with $\sim$3000 AU.
The L$_{\rm clump}$/M$_{\rm clump}$ of IRAS 16547--4247 is 35 L$_{\sun}$/M$_{\sun}$, and it cannot result the decreasing trend of molecular temperature when L$_{\rm clump}$/M$_{\rm clump}$ $>$ 40 L$_{\sun}$/M$_{\sun}$.
Thus the spatial dilution effect is not significant.
Therefore, when L$_{\rm clump}$/M$_{\rm clump}$ $>$ 40 L$_{\sun}$/M$_{\sun}$, the decreasing trend of temperature is not caused by spatial resolution.
The expansion of embedded UC H\,{\footnotesize II} regions might also affect the gas temperature.
The L$_{\rm clump}$/M$_{\rm clump}$ $\sim$ 10 L$_{\sun}$/M$_{\sun}$ is considered a threshold that the birth of the zero age main sequence \citep{2014MNRAS.443.1555U,2016ApJ...826L...8M,2017A&A...603A..33G}, where hydrogen burning begins, then UC H\,{\footnotesize II} regions start to form and disperse the parent clumps \citep{2007prpl.conf..181H}.
Those processes may push ISM into the outer warm envelope, resulting a cavity located in the molecular clouds (e.g., G5.89, see Figure \ref{fig:ContinuumImages}).
Such an expansion of UC H\,{\footnotesize II} region may cause molecular temperature decrease when L$_{\rm clump}$/M$_{\rm clump}$ $\ge$ 40 L$_{\sun}$/M$_{\sun}$, where dissipation may dominate in this phase \citep{2017A&A...603A..33G}.

\subsection{Location of CH$_3$CN-emitting region}
In Figure \ref{fig:ContinuumImages}, the CH$_3$CN emission coincides with the 1.3 mm continuum emission, viewed at our current resolution.
At the same time, we noticed that CH$_3$CN has a relatively complicated morphology in G5.89, where multiple dust continuum regions were found \citep{2008ApJ...680.1271H}.

Assuming that the gas is heated by the central star, molecular temperature is determined by radiative heating, molecular effective distance relative to the central heating source can be estimated by:
\begin{equation}
L = 4\pi R^{2}\sigma T^{4}, \label{eq}
\end{equation}
where $R$ is the distance of the region with temperature $T$, the central source has bolometric luminosity $L$.
We use this formula to estimate molecular effective distance R$_{\rm eff}$, based on the bolometric luminosity of source, and on the molecular temperature derived in Section \ref{sec:MolLine}.
\citet{1987ApJ...318..712K} studied the temperature scales of NH$_3$ gas in the vicinity of UC H\,{\footnotesize II} region G10.6-0.4, where the NH$_3$ gas temperature scales outward with a radius as R$^{-1/2}$.

The molecular effective distances were estimated using eq. \ref{eq} range from $\sim$ 0.003 to $\sim$ 0.083 pc, shown in Table \ref{tab:tab}.
It is likely that CH$_3$CN originated from circum-stellar instead of interstellar.
We also plot the relation between R$_{\rm eff}$ and L$_{\rm clump}$/M$_{\rm clump}$ in the lower left panel of Figure \ref{fig:lm}.
According to our estimation, CH$_3$CN radiation should originate from $\sim$ 1/100 to $\sim$ 1/1000 of the ATLASGAL clump size and there is evidence for a correlation between the R$_{\rm eff}$ and L$_{\rm clump}$/M$_{\rm clump}$ where R$_{\rm eff}$ $\sim$ (L$_{\rm clump}$/M$_{\rm clump}$)$^{0.5\pm0.2}$.
As the star evolves, the molecule is located relatively farther away from the central star.

\subsection{CH$_3$CN abundance as an evolutionary tracer}
In the upper right panel of Figure \ref{fig:lm}, we plot the relation between the molecular abundance and the L$_{\rm clump}$/M$_{\rm clump}$.
It seems that the abundance decreases as L$_{\rm clump}$/M$_{\rm clump}$ increases, where a correlation of X$_{\rm CH_3CN}$ $\sim$ (L$_{\rm clump}$/M$_{\rm clump}$)$^{-1.0\pm0.7}$ can be found.
CH$_3$CN is abundant in the early stage of HMC, and then the abundance decreases with the formation and evolution of UC H\,{\footnotesize II} region.

As stars evolve, they release more and more energy through UV radiation.
Chemical model revealed that molecules have high column density at lower UV fields, while many species are photodissociated away for high UV fields \citep{2004A&A...425..577S}.
CH$_3$CN is subject to such photodissociation by UV radiation.
The species whose abundances are enhanced are simple molecules including radicals and ions in photodissociated region (PDR), there are strong lines of simple species containing 2-4 atoms in W3 IRS4 \citep{1997A&AS..124..205H}, supporting the photodissociation hypothesis.
As shown in the bottom right of Figure \ref{fig:lm}, molecular abundance decreases with the increase of ionization flux (X$_{\rm CH_3CN}$ $\sim$ Ionization flux$^{-0.3\pm0.2}$). It can also support the explanation of the decrease of molecular abundance caused by ionization.

\section{Conclusion} \label{sub:con}
We present SMA observations of 9 massive star-forming regions with L$_{\rm clump}$/M$_{\rm clump}$ ratios ranging from 10 to 154 L$_{\sun}$/M$_{\sun}$.
We detect CH$_3$CN (12$_{\rm K}$--11$_{\rm K}$) lines in all sources, derive molecular temperatures and abundances, and study the relation between CH$_3$CN abundances and the evolutionary stage of sources measured in terms of luminosity-to-mass ratio L$_{\rm clump}$/M$_{\rm clump}$. 

We find that the rotational temperatures of CH$_3$CN increase with the increase of L$_{\rm clump}$/M$_{\rm clump}$ when the ratio is between 10 to 40 L$_{\sun}$/M$_{\sun}$, then seem to decrease when L$_{\rm clump}$/M$_{\rm clump}$ $\ge$ 40 L$_{\sun}$/M$_{\sun}$, where the decline can be explained by dissipation.
Assuming that the CH$_3$CN molecules are heated by radiation from the central stars, we estimated the effective distance of the CH$_3$CN relative to the central heating sources.
Estimated CH$_3$CN effective distance range from $\sim$ 0.003 to $\sim$ 0.083 pc, which accounts for $\sim$ 1/100 to $\sim$ 1/1000 of clump size.
The molecular effective distance R$_{\rm eff}$ increases slightly as L$_{\rm clump}$/M$_{\rm clump}$ increases (R$_{\rm eff}$ $\sim$ (L$_{\rm clump}$/M$_{\rm clump}$)$^{0.5\pm0.2}$).

We also find that the CH$_3$CN abundance is anti-correlated with L$_{\rm clump}$/M$_{\rm clump}$ where X$_{\rm CH_3CN}$ $\sim$ (L$_{\rm clump}$/M$_{\rm clump}$)$^{-1.0\pm0.7}$.
They can be explained by photodissociation.
CH$_3$CN abundance decrease when ionization flux was high.
The strong anti-correlation deserves to be investigated with future high angular-resolution observations, and the relative abundance of CH$_3$CN might serve as a tracer for evolution in future studies.

\begin{sidewaystable*}
\bigskip\bigskip\bigskip\bigskip\bigskip\bigskip\bigskip\bigskip\bigskip\bigskip\bigskip\bigskip\bigskip\bigskip\bigskip\bigskip\bigskip\bigskip\bigskip\bigskip\bigskip\bigskip\bigskip\bigskip\bigskip\bigskip\bigskip\bigskip\bigskip\bigskip\bigskip
\caption{Observational Parameters}
\label{tab:ObsPara}
\centering
\begin{tabular}{lccccccccccc}
\hline\hline
Source & Observation & \multicolumn{2}{c}{Frequency Range}     & Spectral   & \multicolumn{3}{c}{Calibrators} &      & \multicolumn{2}{c}{Synthesized beam} \\
\cline{3-4} \cline{6-8} \cline{10-11}
       & Epoch       & LSB  & USB          & Resolution & Bandpass                        & Gain & Flux          & & FWHM   & P.A. \\
       &             & GHz  & GHz          & MHz        &                                 &      &               & & arcsec & deg  \\
\hline
G5.89--0.39        & 2008 Apr 18           & 219.37--221.34     & 229.37--231.34          & 0.406                & 3C273             & nrao530      & uranus        &            & 2.65 2.18     & +54.0  \\
                   &                       &                    &           &                      &                   & 1921--293    &               &            &               &        \\
G8.68--0.37        & 2008 Sep 17           & 220.28--222.27     & 230.28--232.27          & 0.406                & 3C454             & nrao530      & uranus        &            & 7.86 1.99     & +16.4  \\
                   &                       &                    &           &                      &                   & 1911--201    &               &            &               &        \\
G10.62--0.38       & 2009 Jan 31           & 220.32--222.30     & 230.32--232.30          & 0.406                & 3C454             & 1733--130    & uranus        &            & 5.37 2.66     & +9.2   \\
                   &                       &                    &           &                      &                   & ...          &               &            &               &        \\
G28.20--0.04N      & 2008 Jun 21           & 220.25--222.22     & 230.25--232.22          & 0.406                & 3C454             & 1733--130    & uranus        &            & 2.77 2.59     & +50.8  \\
                   &                       &                    &           &                      &                   & 1911--201    &               &            &               &        \\
G45.07+0.13        & 2007 Apr 13           & 219.46--221.43     & 229.46--231.43          & 0.812                & 3C279             & 1751+096     & callisto      &            & 2.87 1.65     & +77.5  \\
                   &                       &                    &           &                      &                   & 1925+211     &               &            &               &        \\
G45.47+0.05        & 2008 Jun 30           & 219.15--221.13     & 229.15--231.13          & 0.406                & 3C279             & 1830+063     & callisto      &            & 3.11 2.69     & +71.4  \\
                   &                       &                    &           &                      &                   & 1925+211     &               &            &               &        \\
IRAS 16547--4247   & 2006 Jun 06           & 219.21--221.19     & 229.21--231.19          & 0.812                & 3C273             & 1745--290    & uranus        &            & 1.83 0.96     & -7.5   \\
                   &                       &                    &           &                      &                   & 1604--446    &               &            &               &        \\
IRAS 18182--1433   & 2004 Apr 30           & 219.37--221.34     & 229.37--231.34          & 0.812                & 3C273             & nrao530      & uranus        &            & 3.19 2.00     & +16.5  \\
                   &                       &                    &           &                      &                   & 1908--201    &               &            &               &        \\
IRAS 18566+0408    & 2007 Jul 09           & 219.31--221.29     & 229.31--231.29          & 0.812                & 3C273             & 1751+096     & uranus        &            & 3.06 1.48     & +60.6  \\
                   &                       &                    &           &                      &                   & 1830+063     &               &            &               &        \\
\hline
\end{tabular}
\end{sidewaystable*}

\begin{sidewaystable*}[h]
\setlength{\tabcolsep}{2pt}
\bigskip\bigskip\bigskip\bigskip\bigskip\bigskip\bigskip\bigskip\bigskip\bigskip\bigskip\bigskip\bigskip\bigskip\bigskip\bigskip\bigskip\bigskip\bigskip\bigskip\bigskip\bigskip\bigskip\bigskip\bigskip\bigskip\bigskip\bigskip\bigskip\bigskip\bigskip
\caption{Derived parameters}
\label{tab:tab}
\begin{threeparttable}
\begin{tabular}{lcccccccccccccc}
\hline\hline
  &\multicolumn{6}{c}{1.3 mm Continuum Results}  & & \multicolumn{6}{c}{Derived CH$_3$CN parameters} \\
 \cline{2-7} \cline{9-14}
Source  & S$_{\nu}^{\rm Peak}$ & S$_{\nu}^{\rm Total}$ & S$_{\nu}^{\rm Dust}$\tnote{1} & ${\theta}_{\rm s}$\tnote{2} &rms & N$_{\rm H2}$\tnote{3} & &${\theta}_{\rm m}$\tnote{4} &rms & T$_{\rm rot}$ & N$_{\rm tot}$ & X & R$_{\rm eff}$\tnote{5} \\
        & Jy beam$^{-1}$       & Jy                    & Jy                   & arcsec          & mJy beam$^{-1}$            & 10$^{24}$ cm$^{-2}$ & & arcsec              & mJy beam$^{-1}$       & K             & cm$^{-2}$     &   & pc                       \\

\hline
G5.89--0.39       &2.23$\pm$0.07 &7.50$\pm$0.29 &0.91$\pm$0.46 &3.90$\times$3.24 &11.26 &0.70 & &5.4 &39.02 &147$^{+17}_{-12}$ &8.63$^{+1.3}_{-1.0}$(14)  &1.23$\pm$0.78(-9)  &0.016$\pm$0.007 \\
G8.68--0.37       &0.22$\pm$0.01 &0.38$\pm$0.01 &0.25$\pm$0.01 &3.67$\times$2.20 &0.58 &0.39 & &1.2 &15.65 &115$^{+27}_{-17}$ &1.88$^{+0.5}_{-0.6}$(15)  &4.82$\pm$3.69(-9) &0.009$\pm$0.003  \\
G10.62--0.38      &2.98$\pm$0.12 &6.86$\pm$0.37 &3.78$\pm$1.02 &4.33$\times$2.96 &3.21 &2.16 & &2.4 &15.47 &194$^{+49}_{-24}$ &2.51$^{+0.8}_{-0.6}$(15)  &1.16$\pm$0.90(-9) &0.014$\pm$0.006  \\
G28.20--0.04N     &0.87$\pm$0.01 &1.15$\pm$0.01 &0.62$\pm$0.21 &1.73$\times$1.12 &1.50 &1.88 & &1.4 &31.50 &240$^{+46}_{-53}$ &6.23$^{+2.5}_{-3.6}$(15)  &3.31$\pm$3.25(-9) &0.005$\pm$0.002   \\
G45.07+0.13       &1.13$\pm$0.03 &1.63$\pm$0.07 &0.73$\pm$0.42 &1.31$\times$1.18 &4.09 &4.73 & &1.3 &31.14 &143$^{+15}_{-38}$ &2.89$^{+0.3}_{-1.4}$(15)  &6.10$\pm$4.74(-10) &0.026$\pm$0.011 \\
G45.47+0.05       &0.42$\pm$0.01 &0.60$\pm$0.02 &0.35$\pm$0.09 &2.02$\times$1.55 &2.18 &2.15 & &1.5 &30.12 &77$^{+13}_{-6}$   &4.34$^{+1.2}_{-0.9}$(14)  &2.01$\pm$1.47(-10) &0.083$\pm$0.037  \\
IRAS 16547--4247  &0.24$\pm$0.01 &0.56$\pm$0.03 &0.43$\pm$0.07 &1.39$\times$1.25 &19.81 &1.47 & &1.9 &93.34 &237$^{+31}_{-30}$ &2.24$^{+1.3}_{-0.7}$(15)  &1.52$\pm$1.44(-9) &0.003$\pm$0.001 \\
IRAS 18182--1433  &0.28$\pm$0.01 &0.47$\pm$0.01 &0.46$\pm$0.01 &2.37$\times$1.17 &1.09 &1.64 & &1.4 &14.22 &145$^{+16}_{-22}$ &1.12$^{+0.3}_{-0.2}$(15)  &6.82$\pm$4.63(-10) &0.005$\pm$0.002  \\
IRAS 18566+0408   &0.16$\pm$0.01 &0.31$\pm$0.02 &0.30$\pm$0.01 &2.10$\times$1.80 &2.08 &0.58 & &1.2 &28.85 &193$^{+37}_{-25}$ &7.25$^{+1.4}_{-2.9}$(15)  &1.25$\pm$0.97(-8) &0.003$\pm$0.001   \\
\hline
\end{tabular}
\begin{tablenotes}
\item[1] S$_{\nu}^{\rm Dust}$ = S$_{\nu}^{\rm Total}$ -- S$_{\nu}^{\rm free-free}$. The free-free emission was obtained from \citet{2014ApJ...786...38H}, who assumed the free-free emission is optically thin from 10 to 100 GHz and estimated the free-free emission by $S_{\nu}\propto \nu^{-0.1}$.
\item[2] Deconvolved continuum sizes from 2D gaussian fit.
\item[3] H$_2$ column density derived from the estimated 1.3 mm dust continuum emission (S$_{\nu}^{\rm dust}$) assuming T$_{\rm dust}$ = T$_{\rm CH_3CN}$.
\item[4] Deconvolved molecular distribution sizes from 2D gaussian fit.
\item[5] Effective distance of the CH$_3$CN-emitting region is calculated using $L = 4\pi R^{2}\sigma T^{4}$. $T$ is the temperature of the CH$_3$CN gas and $L$ is clump bolometric luminosity.
\end{tablenotes}
\end{threeparttable}
\end{sidewaystable*}

\bibliographystyle{raa}
\bibliography{bibtex}

\end{document}